\documentclass[aps,pra,nofootinbib,preprint,amsmath,amssymb,floatfix]{revtex4}
\usepackage{color}
\usepackage{graphicx}

\begin{document}

\newcommand{\tc}{\textcolor}
\newcommand{\g}{blue}
\newcommand{\ve}{\varepsilon}
\title{Rip brane cosmology from a viscous holographic dark fluid}         

\author{  I. Brevik$^1$  }      
\affiliation{$^1$Department of Energy and Process Engineering,  Norwegian University of Science and Technology, N-7491 Trondheim, Norway}
\author{A. V. Timoshkin$^{2,3}$}
\affiliation{$^2$Tomsk State Pedagogical University, Kievskaja Street, 60, 634061 Tomsk, Russia}
\affiliation{$^3$Tomsk State University of Control Systems and Radio Electronics, Lenin Avenue, 36, 634050 Tomsk, Russia}

\date{\today}          

\begin{abstract}
This article is devoted to the application of the holographic principle to  describe Rip brane cosmological models in the presence of a bulk viscosity. We make use of the generalized infrared-cutoff holographic dark energy, introduced by Nojiri and Odintsov.       We consider various examples: Rip brane cosmology corresponding to the Little Rip case,   asymptotic de Sitter theory, and the so-called  Big Freeze theory leading to a singularity. Analytical expressions for infrared cutoffs, as well as the particle and the future horizons at the brane, are obtained. The equations for energy conservation on the brane within the holographic theory are obtained in each case. The correspondence between viscous cosmology   and holographic cosmology on the brane is shown.

\end{abstract}
\maketitle
Keywords: Viscous cosmology; rip brane cosmology; holographic dark fluid

Mathematics Subject Classification 2010: 83F05
\bigskip
\section{Introduction}
	
An interesting approach to the explanation of the accelerated expansion of the Universe in terms of the dark energy concept, is the application of the so-called holographic principle \cite{1}.  The generalized cutoff holographic dark energy model was proposed by Nojiri and Odintsov  \cite{2,3}. Further investigations with application to holographic dark energy in the late-time cosmology were presented in several papers \cite{4,5,6,7,8,9,10,11,12,13}. The holographic principle can be realized also in the early universe, obtaining inflation \cite{14},  and a cosmological bounce \cite{15},  having a  holographic origin. It has been  shown that the application of the theory of holographic dark energy in the late-time universe is consistent with astronomical observations \cite{16,17,18,19,20}.  General reviews of different approaches to dark energy are given in Refs.~\cite{21} and \cite{22}.

In the late-time universe there is a  possibility of forming a  brane cosmological model \cite{23,24}.  The brane scenario is one among several approaches for how to solve the singularity  problem \cite{25}. Several dark energy models on the brane, with various  scenarios of evolution, have been investigated in Refs.~\cite{26,27,28,29}.

The present article is devoted to an  application of the principle of a generalized holographic cutoff \cite{3} in order to describe    the brane dark energy models. We will start from the basic brane models  proposed in \cite{26,27},  and investigate the evolution of the cosmic fluid assuming that the fluid possesses a bulk viscosity. The most convenient formalism for constructing  the (scalar or fluid) dark energy is to use the  equation of state (EoS). The evolution of the universe then depends on the choice of the EoS. In accordance with usual 4d Friedman-Robertson-Walker (FRW) cosmology, we will consider the Little Rip behavior, the asymptotic de Sitter evolution,  and the Big Freeze singularity. The viscous contribution is effectively taken into account via the corresponding choice of an inhomogeneous dark fluid. We will apply the holographic principle to the late-time epoch, associating the choice made by  Nojiri and Odintsov for the infrared cutoff with future events and particle horizons, and we will derive the corresponding forms of the energy conservation law. Thereby we establish the equivalence between our viscous fluid models on the brane  and the holographic models on the brane.

	In the next section we recall the main points of the holographic principle applied to the description of the  universe. The third section is devoted to the study of cosmological models on the brane, taking into account the viscosity property of the dark fluid, showing the equivalence between viscous fluid cosmology  and holographic fluid cosmology on the brane. The final section summarizes the obtained results.

\section{The holographic dark energy}

In this section we give the main points of the holographic principle, following the terminology proposed by Li \cite{17} (string theory uses a different terminology \cite{30}).
The holographic description of dark energy, based on the holographic principle, states that  the cutoff radius of the horizon may be related to the infrared cutoff. The generalized holographic energy density was proposed in Ref.~\cite{2},  where the infrared cut-off was identified with a combination of FRW universe parameters: the Hubble parameter, the particle and the future event horizons, the cosmological constant, and the finite life time of the universe.

In general the holographic energy density is proportional to the inverse squared infrared cutoff   $L_{IR}$ \cite{3}
\begin{equation}
\rho=\frac{3c^2}{k^2L_{IR}^2}, \label{1}
\end{equation}			
where  $k^2=8\pi G$ is  Einstein’s gravitational constant and $c$   is a dimensionless parameter (thus not the light velocity).

In order to apply  equation (\ref{1}) in a cosmological framework we consider a homogeneous and isotropic FRW flat universe
\begin{equation}
ds^2=-dt^2+a^2(t)\sum_{i=1,2,3} (dx^i)^2, \label{2}
\end{equation}
where $a$  is the scale factor.
The first FRW equation has the form
\begin{equation}
H^2=\frac{k^2}{3}\rho, \label{3}
\end{equation}
where  $\rho$ is the holographic energy density. Then,
\begin{equation}
H=\frac{c}{L_{IR}}. \label{4}
\end{equation}
Actually there is no definite prediction for the choice of the infrared cutoff at present; one can choose    the particle horizon,   or the future event horizon, whose definitions are given in Ref.~(\ref{3}),
\begin{equation}
L_p \equiv a(t)\int_0^t\frac{dt'}{a(t')}, \quad L_f(t) \equiv a(t)\int_t^\infty \frac{dt'}{a(t')}. \label{5}
\end{equation}
In the  general case one can construct an infrared cutoff as a combination of these quantities and their derivatives \cite{2}. Using the infrared radius in this form, one can apply the holographic principle to describe the late-time universe.

Further, we will assume that the viscous fluid,  responsible for the accelerated expansion of the universe, is related to  the holographic energy density.

\section{Holographic description of rip brane  FRW viscous cosmology}

One of several ways to solve the singularity problem is to apply the brane world scenario. According to the simplest version the space-time is homogeneous and isotropic along the three spatial dimensions, this being our 4-dimension universe,  while there is a thin wall and constant spatial curvature when viewed from  a 5-dimensional space-time \cite{31,32,33}.  The Friedmann equation on the brane turns out to be  different from the standard cosmology equation. The Hubble parameter has the following form
\begin{equation}
H^2=\frac{k^2}{3}\rho \left( 1+\frac{\rho}{2\lambda}\right), \label{6}
\end{equation}
where $H=\dot{a}/a$ and $\lambda$ is the brane tension. If the inequality $\rho \ll  |\lambda|$ holds, then Eq.~(\ref{6}) differs insignificantly from the first FRW Eq.~(\ref{3}). Thus, one can assume that in our epoch  $\rho/2\lambda \ll 1$, and there will be no significant difference between the brane model and conventional FRW cosmology.

Next, we consider examples of  holographic dark energy models on the brane, corresponding to  the Little Rip model, the de Sitter asymptotic theory, and the so-called  type III cosmological singularity model. We take   into account the bulk viscosity of the fluid throughout.  We will suppose, for simplicity, that the universe consists of dark energy only.
We assume that this universe  satisfies an inhomogeneous equation of state (EoS) in flat FRW space-time \cite{34,35},
\begin{equation}
p=\omega(\rho,t)\rho -3H\zeta (H,t), \label{7}
\end{equation}
where  $\omega(\rho,t)$  is the thermodynamic parameter and  $\zeta (H,t)$   the bulk viscosity, depending  on the Hubble parameter  and  on the time $t$. According to conventional thermodynamic, it is natural to  assume that $\zeta >0$.
We will take the following form for the thermodynamic EoS parameter  \cite{34,35}
\begin{equation}
\omega(\rho,t)=\omega_1(t)(A_0\rho^{\alpha-1}-1), \label{8}
\end{equation}
 where $A_0 \neq 0$ and
$\alpha \geq 1$ are constants.  We choose the bulk viscosity as \cite{34,35}
\begin{equation}
\zeta(H,t)=\zeta_1(t)(3H)^n, \label{9}
\end{equation}
with $n >0$.

As mentioned, we assume that the universe consists of a one-component viscous fluid, whose energy conservation equation is
\begin{equation}
\dot{\rho} +3H(\rho+p)=0. \label{10}
\end{equation}
Further, we will investigate the future evolution of the universe on the brane, in the holographic picture. For this we will apply the holographic principle  on the brane for different cases, and for different forms for the bulk viscosity term,    in order to obtain the appropriate  energy conservation law in each case.

\subsection{Little Rip case}

Let us consider a nonsingular brane Little Rip model, in which the scale factor is given as \cite{26,27}
\begin{equation}
a(t)=a_0\exp \left[ \frac{\lambda}{3\alpha^2}\cosh \left( \sqrt{\frac{3\alpha^3}{2\lambda}}\, t\right)\right], \quad \lambda >0. \label{11}
\end{equation}
Here it is natural to associate $t=0$ with the present time; then $a_0\exp(\lambda/3\alpha^2)$ corresponds to the present time scale factor. Various scenarios of the Little Rip were discussed in Refs.~\cite{36,37}.

We now adopt a constant value for the EoS parameter, $\omega(\rho,t)=\omega_0$, and assume a linear dependence of the bulk viscosity on the Hubble parameter, $\zeta (H,t)=3\tau H$, with $\tau$ a positive dimensional constant. Then taking into account Eq.~(\ref{6}), we obtain the EoS (\ref{7}) in the form
\begin{equation}
p=\omega_0\left( \sqrt{\lambda^2+6\frac{\lambda}{k^2}H^2}-\lambda \right) -9\tau H^2. \label{12}
\end{equation}

Let us choose to consider the holographic principle on the brane in terms of the future horizon $L_f$, \cite{5}. After some calculation we obtain
\begin{equation}
L_f=\sqrt{\frac{\lambda}{6\alpha^3}}\left\{ \begin{array}{ll}
\exp \left[ \frac{\lambda}{3\alpha^2}\left( \cosh \sqrt{\frac{3\alpha^3}{2\lambda}}\, t+\frac{1}{\sqrt{2}}\right) \right]
 E_1\left[ \cosh \sqrt{\frac{3\alpha^3}{2\lambda}}\, t+\frac{\lambda}{3\sqrt{2}\alpha^2}\right] + \\
+\exp \left[ \frac{\lambda}{3\alpha^2} \left(\cosh \sqrt{ \frac{3\alpha^3}{2\lambda}}\, t-\frac{1}{\sqrt{2}}\right) \right]
E_1\left[ \cosh \sqrt{\frac{3\alpha^3}{2\lambda}}\, t-\frac{\lambda}{3\sqrt{2}\alpha^2}\right],
\end{array}
\right\}
\end{equation}
where $E_1(x)$ is the integral exponential function.

The expressions for the Hubble parameter in terms of the future event horizon $L_f$ and its time derivatives are \cite{3}
\begin{equation}
H=\frac{\dot{L}_f+1}{L_f}, \quad \dot{H}=\frac{\ddot{L}_f}{L_f}-\frac{\dot{L}_f^2}{L_f^2}-\frac{\dot{L}_f}{L_f^2}. \label{14}
\end{equation}
Thus by using Eqs.~(\ref{12}) and (\ref{14}) the energy conservation equation (\ref{10}) takes the following form in the holographic language,
\begin{equation}
\frac{2\lambda}{k}\, \frac{\ddot{L}_f-\frac{\dot{L}_f^2+\dot{L}_f}{L_f}}{\sqrt{(\lambda kL_f)^2+6\lambda (\dot{L}_f+1)^2}}-9\tau \left( \frac{\dot{L}_f+1}{L_f}\right)^2 +
\end{equation}
\[ +(\omega_0+1)\left[ -\lambda +\frac{1}{kL_f}\sqrt{ (\lambda kL_f)^2+6\lambda (\dot{L}_f+1)^2}\right]=0. \]
This is thus he realization of the holographic principle on the brane for a viscous fluid within the Little Rip model.

\subsection{Asymptotic de Sitter regime}

In this case the brane has a negative tension  $(\lambda <0)$, so that  the universe expands in a quasi-de Sitter regime.
The scale factor increases with time as \cite{26,27}
\begin{equation}
a(t)=a_0\exp \left\{ -\frac{\beta^2}{2}\left[ \frac{1}{\cos \eta_0}-
\sqrt{ 1+\left(\tan \eta_0+\frac{\alpha}{\beta}t \right)^2} \right]\right\}. \label{16}
\end{equation}
Here  $\beta^2=\frac{2|\lambda|}{3\alpha^2}$ is a dimensionless  parameter and $\eta_0=\sqrt{\frac{
3}{2\lambda}}\,\alpha^2t_0$,  where $t_0$ is the present time.

Let us study the case of constant bulk viscosity, $\zeta(H,t)=\zeta_0 >0$, and assume that    the thermodynamic parameter $\omega(\rho,t)$
 is a linear function of the energy density,
 \begin{equation}
 \omega(\rho,t)=A_0\rho -1. \label{17}
 \end{equation}
The the EoS for the viscous fluid (\ref{7}) takes the form
\begin{equation}
p=(2A_0\lambda +1)\left( \lambda -\sqrt{\lambda^2+\frac{6\lambda}{k^2}H^2}\right) +3H\left( \frac{2A_0\lambda}{k^2}H+\xi_0\right). \label{18}
\end{equation}
The holographic description of a viscous dark fluid on the brane, in terms of the future event horizon $L_f$, Eq.~(\ref{5}), then becomes
\begin{equation}
L_f=\frac{\beta}{\alpha}\left\{ \begin{array}{ll}
\frac{2}{\beta^2}-\left( \frac{\alpha}{\beta}t+\tan \eta_0\right) +\sqrt{1+\left( \frac{\alpha}{\beta}t+\tan \eta_0\right)^2} +  \\
+\frac{\beta^2}{4}\exp \left[ \frac{\beta^2}{2}\sqrt{ 1+\left( \frac{\alpha}{\beta}t+\tan \eta_0\right)^2}\right] E_i\left[ -\frac{\beta^2}{2}\sqrt{ 1+\left(\frac{\alpha}{\beta}t+\tan \eta_0\right)^2}\right]
\end{array}
\right\}, \label{19}
\end{equation}
where $E_i$ is an exponential integral function.

 Now using Eqs.~(\ref{14}) and (\ref{18}) we can  rewrite the continuity equation for the energy in a holographic language,
\begin{equation}
\frac{\ddot{L}_f-\frac{\dot{L}_f}{L_f}(\dot{L}_f+1)}{k\sqrt{(\lambda kL_f)^2+6\lambda (\dot{L}_f+1)^2}} -\frac{3\xi_0}{2\lambda}\, \frac{\dot{L}_f+1}{L_f}+
\end{equation}
\[ +A_0\left[\lambda -\frac{1}{kL_f}\sqrt{ (\lambda kL_f)^2+6\lambda (\dot{L}_f+1)^2} +\frac{3}{(kL_f)^2}(\dot{L}_f+1)^2\right]=0. \]

Thus, we have constructed a holographic model on the  brane, taking into account the fluid viscosity in the de Sitter asymptotic regime.

\subsection{	Big Freeze singularity cosmology}

Let us assume that the brane tension is positive, $\lambda >0$.  There are then two Big Freeze singularities: one in the past ($t \rightarrow -\infty$ ) and the other  in the future ($t\rightarrow \infty$). The universe begins its existence at  $t_{in}=-\frac{1}{\alpha}\sqrt{\frac{2\lambda}{3\alpha}}$  and ends at $ t_f=+\frac{1}{\alpha}\sqrt{\frac{2\lambda}{3\alpha}}$.

The scale factor is \cite{26,27}
\begin{equation}
a(t)= a_f\exp \left[-\frac{\lambda}{3\alpha^2}\left( 1-\frac{3\alpha^3t^2}{2\lambda}\right)^{1/2}\right], \label{21}
\end{equation}
where $a_f$ is the final scale factor. This expression shows that the universe contracts in the time interval $t_{in}<t<0$ and then expands.

In this model we will assume the thermodynamic parameter $\omega(\rho,t)$  to be  a linear function of the energy density (\ref{17}), and also assume the bulk viscosity $\zeta(H,t)$ to be a linear function  of $H$.
Then, following (\ref{7}, we write the EoS as
\begin{equation}
p=2A_0\lambda^2-(2A_0\lambda +1)\sqrt{ \lambda^2+6\frac{\lambda}{k^2}H^2} +3\left( \frac{2A_0\lambda}{k^2}-3\tau\right)H^2, \label{22}
\end{equation}
and calculate the particle horizon, following (\ref{5}), as
\begin{equation}
L_p=2\sqrt{\frac{\pi}{\alpha}}\exp \left[ \frac{\lambda}{3\alpha^2}\left( 1-\sqrt{ 1-\frac{3\alpha^3}{2\lambda}t^2}\right) \right] {\rm erf} \left( \frac{1}{2}\sqrt{\alpha}\, t\right), \label{23}
\end{equation}
where ${\rm erf}\left( \frac{1}{2}\sqrt{\alpha}\, t\right)$   is a probability integral.

Substituting expressions for the Hubble parameter in terms of the particle horizon  $L_p$ and its time derivatives \cite{3}, we get
\begin{equation}
H=\frac{\dot{L}_p-1}{L_p}, \quad \dot{H}=\frac{\ddot{L}_p}{L_p}-\frac{\dot{L}_p^2}{L_p^2}+\frac{\dot{L}_p}{L_p^2}. \label{24}
\end{equation}	
Moreover, substituting the EoS in the form (\ref{22}) in 	the continuity equation for energy (\ref{10}), we get in this cosmological model
\begin{equation}
\frac{\ddot{L}_p-\frac{\dot{L}_p}{L_p}(\dot{L}_p-1)}{k\sqrt{(\lambda kL_p)^2+6\lambda (\dot{L}_p-1)^2}}-3\left( \frac{A_0}{k^2}-\frac{3\tau}{2\lambda}\right)\left( \frac{\dot{L}_p-1}{L_p}\right)^2+ \label{25}
\end{equation}
\[ +A_0\left[ \lambda -\frac{1}{kL_p}\sqrt{(\lambda kL_p)^2+6\lambda (\dot{L}_p-1)^2}\right]=0. \]
This shows the behavior of the  Big Freeze singularity within the  viscous holographic model.

\section{Conclusion}

In the present work we have applied the holographic principle to late-time cosmology on the brane, considering various  cosmological models for the rip brane holographic scenario. The mathematical tool for description was the inhomogeneous equation of state for a one-component dark fluid having a bulk  viscosity. For each of the considered models,  using analytical calculations, we arrived at  expressions for the infrared cutoff in the form of a particle horizon, or an event horizon.  We reformulated the continuity equation for the energy density of the viscous fluid in the holographic language. Thereby, the equivalence between the description of cosmology on the brane using a viscous fluid, and the holographic description using infrared cutoff as proposed by Nojiri and Odintsov \cite{2,3},  was shown. This theory can be extended to the case two coupled fluids with calculations similar to those given here.

\bigskip

A natural question is whether the predictions of the   holographic theory agree with  astronomical observations. An analysis of the correspondence between theoretical models of holographic dark energy on the brane and  astronomical observations  was carried out in \cite{38}, whereby  it became possible to determine the upper limit of the  ratio between the present  energy density and tension on the brane. In that article the relationships between apparent magnitudes and redshifts for distant supernovae Ia, Hubble parameters for various redhifts, and baryon acoustic oscillations were obtained. For a  wide range of parameters, the observational  data were well described.

\end{document}